# PECULIAR VELOCITIES OF 3000 SPIRAL GALAXIES FROM THE 2MFGC CATALOG

I.D.Karachentsev[1], Yu.N. Kudrya[2], V.E. Karachentseva[2], and S.N. Mitronova[1]

[1] Special Astrophysical Observatory, Russian Academy of Sciences, Russia; e-mail: ikar@sao.ru
[2] Astronomical Observatory of the Taras Shevchenko Kiev National University, Ukraine

The 2MFGC catalog we have used contains 18020 galaxies selected from the extended objects in the 2MASS infrared sky survey as having apparent ratios of the axes $b/a < 0.3$. Most of them are spiral galaxies of late morphological types whose discs are seen almost edge-on. The individual distances to the 2724 2MFGC galaxies with known rotation velocities and radial velocities are determined using a multi-parametric infrared Tully-Fisher relation. A list of the distances and peculiar velocities of these galaxies is presented. The bulk motion of the 2MFGC galaxies relative to the cosmic microwave background is characterized with the velocity $V = 199\pm37$ km/s in the direction $l = 304°\pm11°$, $b = -8°\pm8°$. Our list is currently the most representative and uniform sample for analyzing non-Hubble motion of galaxies on a scale of ~100 Mpc.

Keywords: *galaxies: Tully-Fisher relation: large-scale motions*

## 1. Introduction

According to the conventional definition, the redshift of galaxies is the sum of two terms: the isotropic cosmic expansion velocity and the peculiar velocity owing to gravitational attraction by the surrounding matter. In practice, determining the peculiar velocity of a galaxy requires knowledge of both its observable radial velocity relative to some reference system and the distance to the galaxy determined independently of the radial velocity. The most widely used method of determining the individual distances to spiral galaxies is the Tully-Fisher method (TF) [1]. Peebles [2] has shown that in terms of a linear theory of gravitational instability, the peculiar velocities of galaxies are related to fluctuations in the mass through the cosmological parameter. Thus, the observed distribution of peculiar radial velocities of galaxies within a sufficiently large volume can be used to recover the distribution of matter within that volume, given a set of cosmological parameters and boundary conditions [3].

Solving this sort of problem requires extensive, uniform samples of galaxies which are evenly and sufficiently densely distributed over the sky and have great depth. This condition is satisfied by the RFGC [4], a catalog of flat spiral galaxies which are visible edge-on. The RFGC contains 4236 galaxies with an optical angular diameter $a \geq 0'.6$ and an apparent ratio of the axes, $a/b \geq 7$. In 2000, the radial velocities and rotation amplitudes (either $W_{50}$ or $W_{20}$ for an HI line, or the optical $V_{max}$) had been measured for only 1327 of the RFGC galaxies (31% of the entire catalog). A multiparameter TF relation was constructed for them based on the cataloged characteristics of the galaxies and their peculiar distances were determined [5]. The accuracy of the velocities determination was roughly 20-25%. These data have been used to estimate the parameters of the bulk motion and to construct the two-dimensional velocity field [6]. Feldman et al. [7] have compared the dependence of the differences of the velocities of paired galaxies on their mutual distance for different samples: MarkIII [8], SFI [9,10], ENEAR [11], and RFGC. It was shown that, although these catalogs differ in sample size and geometry, the results for the cosmological parameter of the density of matter and the amplitude of the fluctuations in the density are stable and in fairly good agreement.

Unfortunately, data on the rotational velocities of RFGC galaxies in both the radio and optical ranges accumulates rather slowly. Since the publication of the list [5], the number of galaxies with known radial velocities and rotation amplitudes has increased by roughly a quarter.

Beginning with the first papers of Aaronson et al. [12-14], researchers have made extensive use of the infrared TF relation (IRTF). Its advantages have been described in detail by the review by Strauss and Willick [3]. We have constructed the IRTF relation in the *J, H*, and $K_S$ bands for RFGC galaxies [15] using photometric characteristics from the 2MASS Extended Source Catalog, XSC [16]. Then, taking the Kron $J_{fe}$ value as a reference and constructing a linear multi-parameter IRTF relation, we determined the amplitude and apex of the bulk motion: *V = 199±61* km/s, *l* = 301° ± 18°, and *b* = −2° ± 15° for a sample of 971 galaxies with velocities within 18000 km/s [17]. This result is in good agreement with other published data [18].

## 2. A new sample and Tully-Fisher relation

The successful application of the IRTF relation to the RFGC galaxies has revealed a new possibility for sampling flattened spiral galaxies from the XSC catalog based on their 2MASS characteristics. The new 2MASS selected Flat Galaxy Catalog, 2MFGC [19] contains 18020 galaxies with infrared axis ratios *b/a < 0.3*, which corresponds roughly to an optical axis ratio *a/b> 6*. A statistical analysis of the characteristics of the 2MFGC galaxies shows that, despite the poor sensitivity of the 2MASS sample to galaxies with a low surface brightness and blue dwarf systems, the 2MFGC catalog is an entirely adequate "optical" RFGC catalog: the new catalog contains mainly spiral galaxies of later types. The advantage of the 2MFGC catalog compared to the RFGC is the four times larger sample size, as well as much greater depth and more completeness of the data in the region of the Milky Way. It is also important that the photometric parameters of the 2MASS galaxies have been measured in a uniform way for the overwhelming majority of the objects (except those with the largest angular diameters).

For compiling a summary of the radial velocities $V_h$, widths of the hydrogen lines $W_{20}$ and $W_{50}$, and the optical estimates of the rotational amplitude $V_{max}$, we have used the following sources:
(1) The latest versions of the LEDA and NED data bases.
(2) A list of RFGC galaxies with peculiar velocities [5].
(3) A list of southern RFGC galaxies identified in the HIPASS survey [20] and in the catalog of Kilborn et al. [21].
(4) Data from our observations on the 100-m Effelsberg telescope [22,23].
(5) A summary catalog of the profiles of the hydrogen 21 cm line in optically selected galaxies [24].
(6) Spectra of the $\alpha$ line obtained on the 5-m Palomar telescope [15].

In all, our data set included about 5700 2MFGC galaxies with measured radial velocities, along with 3110 galaxies for which estimates of the rotational velocity were also available. We then eliminated cases of unreliable photometry (adjacency to a bright star), low signal/noise ratios for HI line profiles, and possible effects of near neighbors on the HI line profile from the sample, and also excluded galaxies with peculiar morphology. This left 5633 galaxies with measured $V_h$, of which 3074 have measured rotation velocities; this represents 31% and 17%, respectively, of the total number of galaxies in the 2MFGC catalog.

Compared to the sample used in our previous paper [26], there are 309 more galaxies with measured rotation velocities.

In the earlier work [26] we provided a detailed justification for the use of the hydrogen line width $W_{50}$ as the basic width. Here we use the previously found relations for converting $W_{20}$ and $V_{max}$ to $W_{50}$: $W_{50} = W_{20} - (29.5 \pm 0.5)$ km/s, and $W_{50} = 2V_{max} + (27.4 \pm 7.8)$ km/s.

The calculations were carried out in two steps. In the first, we calibrate the linear multiparameter IRTF relation, taking it in the form

$$M = C_0 + C_1 \cdot \log W^c_{50} + C_2 \cdot \log(a/b) + C_3 \cdot Jhl + C_4 \cdot (J^c_{fe} - K^c_{fe}) + C_5 \cdot Jcdex , \qquad (1)$$

where $W^c_{50}$ is the width $W_{50}$ of the hydrogen line corrected for cosmological broadening, $Jhl$ is the effective surface brightness in the $J$ band, $Jcdex$ is the concentration index (the ratio of the radii within which 3/4 and 1/4 of the galaxy light is concentrated), $J^c_{fe} \equiv J_{fe} - A_J$ and $K^c_{fe} \equiv K_{fe} - A_K$ are the Kron $J$- and $K$-magnitudes corrected [27] for extinction in our galaxy (we use the difference in magnitudes as a color index), and $a/b$ is the ratio of the axes.

We calculated the absolute magnitude from the apparent one $J^c_{fe}$ in the usual way:

$$M = J^c_{fe} - 25 - 5 \log r. \qquad (2)$$

For calibrating Eq. (1) we estimated the photometric distance $r$ (in Mpc) using the post-Hubble relation

$$r = V_{3K} \{1 - (q_0 - 1)V_{3K}/2c\}/H_0 , \qquad (3)$$

which is valid for a uniform, isotropic cosmological model. (Here $c$ is the speed of light.) The radial velocity $V_{3K}$ in the microwave background system was calculated from the heliocentric radial velocity $V_h \equiv cz$ using the parameters of the Sun's motion relative to the microwave background [28]. The Hubble constant was taken to be $H_0 = 75$ km/s/Mpc. The value $q_0 = -0.55$ of deceleration parameter corresponds to the standard cosmological model with cold dark matter and the cosmological constant ($\Omega_m = 0.3$, $\Omega_\lambda = 0.7$). We find the coefficients in Eq. (1) by minimizing the sum of the squares of the deviations in the right hands sides of Eqs. (1) and (2) taking Eq. (3) into account.

In the second step, after calibration of the IRTF relation, we find the distance according to Eqs. (1) and (2) and calculate the individual peculiar velocity of a galaxy in the post-Hubble approximation:

$$V_{pec} = V_{3K} - H_0 r \{1 + (q_0 - 1)H_0 r/2c\}, \qquad (4)$$

For the sample with $N = 3074$ we obtained a rather large scatter of the points in the IRTF diagram, $\sigma_{TF} = 0^m.76$. As in Ref. 26, we eliminated galaxies which deviate by more than $3\sigma_{TF}$ in the IRTF diagram, as well as those galaxies whose individual peculiar velocities exceed 3000 km/s in the 3K-system, assuming that these deviations are caused by measurement errors rather than physical causes. After several iterations of the exclusion process (until the process converged), we obtained a sample with $N = 2724$ galaxies, which was characterized by standard deviation $\sigma_{TF} = 0^m.47$.

The coefficients in the IRTF relation (1) together with their significance according to the Fisher criterion (given in parentheses) turned out to be:

$$C_0 = -9.95 \pm 0.43 \text{ (541)}, \quad C_1 = -6.53 \pm 0.08 \text{ (6696)}, \quad C_2 = 1.17 \pm 0.05 \text{ (466)},$$
$$C_3 = 0.228 \pm 0.016 \text{ (208)}, \quad C_4 = -0.53 \pm 0.07 \text{ (53)}, \quad C_5 = -0.016 \pm 0.009 \text{ (3)}.$$

We now comment briefly on the role of the coefficients in the regression results.

*log(a/b)*: the positive value of the coefficient $C_2$ means that in most flat/inclined galaxies the luminosity is fainter than the average;

*Jhl*: the positive value of $C_3$ means that in the more diffuse galaxies the luminosity is weaker (the periphery is underestimated);

$J^c_{fe} - K^c_{fe}$ - : the negative value of $C_4$ means that in the redder galaxies, where the contribution of the disk is smaller, the luminosity is higher than the average (the losses at the edge are low);

*Jcdex*: the coefficient $C_5$ may be essentially insignificant because the concentration index *Jcdex* is correlated with *Jhl* and/or with $J^c_{fe} - K^c_{fe}$.

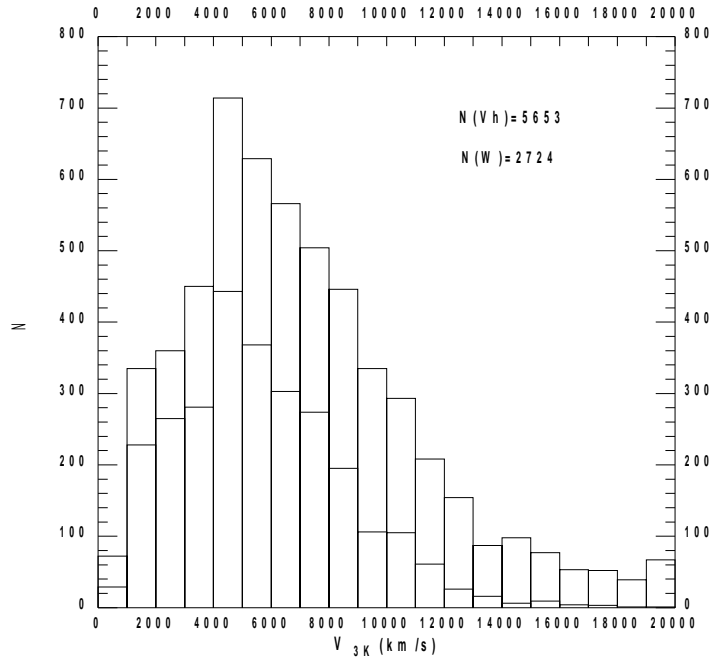

Fig. 1. The distributions with respect to $V_{3K}$ for $N = 5539$ sample galaxies with radial velocities $V_{3K} < 20000$ km/s (thin lines) and a subsample of $N = 2724$ galaxies for which the peculiar velocities are given (thick lines).

Here two comments are appropriate. (1) By eliminating 11% of the galaxies from the sample, we reduced the scatter in the IRTF diagram by a factor of 1.6. An additional check of the excluded galaxies shows that using new, more accurate data generally eliminates the strong deviations. (2) The 2MFGC did not include about 600 RFGC galaxies with known $V_h$, $W$, and $V_{max}$, since they did not meet the sample criteria for one or another reason: of them 188 have 2MASS photometry. In principle, these data could also be included in the analysis, but for the sake of uniformity we have not combined the 2MASS and optical characteristics.

The 2MFGC galaxies with measured radial velocities (N = 5653) cover a broad range of $V_{3K}$ up to 52300 km/s with an average value of 7450 km/s. Figure 1 shows the distribution of N = 5539 2MFGC galaxies from this sample with respect to the radial velocities in the interval 0-20000 km/s (thin lines). The galaxies with measured rotation amplitudes, for which we give the peculiar velocities (N = 2724), are closer: the maximum velocity in this sample is $V_{3K}$ = 19100 km/s and the average velocity is $V_{3K}$ = 5670 km/s. The thick lines in Fig. 1 indicate the distribution of the galaxies in this subsample.

Figure 2 shows the distributions of both samples with respect to the corrected apparent magnitude: N = 5653 galaxies with measured radial velocities (thin lines) and N = 2724 galaxies for which we give the peculiar velocities. These figures show that the galaxies with measured rotational amplitudes represent more than half of the galaxies with $V_{3K}$ < 8000 km/s and of the objects brighter than $J^c_{fe}$ = 13$^m$.

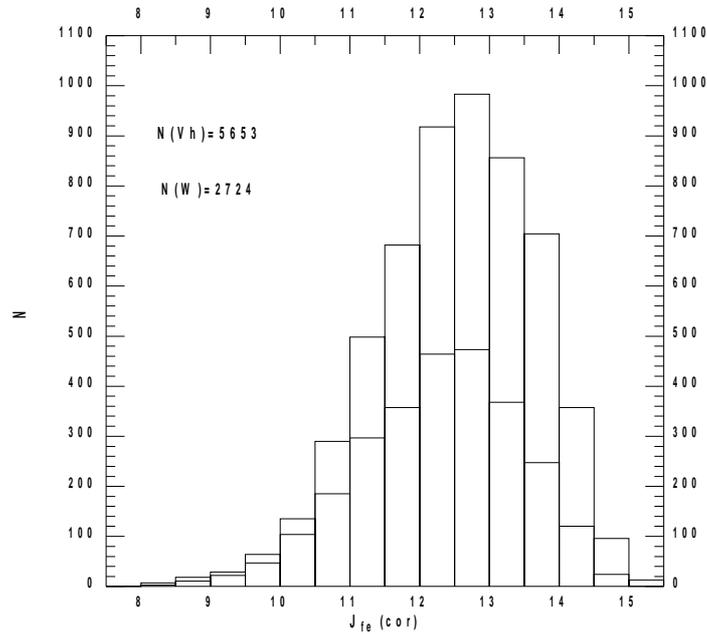

Fig. 2. The distributions with respect to the corrected magnitude $J^c_{fe}$ for the sample of N = 5539 galaxies with radial velocities (thin lines) and for the subsample of N=2724 galaxies for which the peculiar velocities have been calculated (thick lines).

The distribution over the sky of the 2724 galaxies from Table 1 is shown in Fig. 3 in galactic coordinates.

## 3. List of the peculiar velocities and characteristics of the bulk motion

Table 1 is a list of the peculiar velocities of some of the 2724 2MFGC galaxies.* The column entries are: (1) the number in the 2MFGC catalog; (2) the galactic coordinates (in degrees); (3) $J^c_{fe}$, the Kron "elliptical" stellar magnitude, corrected for absorption in our galaxy; (4) $W_{50}$, the width of the 21 cm line, corrected for cosmological broadening (the width $W_{20}$ and rotation amplitude $V_{max}$ are reduced to $W_{50}$ using the equations given in Section 2); (5) $V_h$, the heliocentric radial velocity;

---

* Here we present the beginning of the table as an illustration. The complete list of peculiar velocities will be provided in the astrophysical data base.

(6) $V_{3K}$, the radial velocity in the 3K-system calculated according to Ref. 28; (7) $H_0 r$ the photometric distance obtained using Eqs. (1) and (2); (8) $V_d$, the dipole component of the individual radial velocity of the galaxy (the part of the radial velocity of the galaxy determined by the dipole bulk motion); (9) $V_{pec}$, the peculiar velocity of the galaxy calculated in the post-Hubble approximation according to Eq. (4); and, (10) $V_p = V_{pec} - V_d$, the noise component of the peculiar velocity (this has the same significance as in column (12) of the list of peculiar velocities in Ref. 5). The units in columns 4-10 are km/s.

TABLE 1. Catalog of Peculiar Velocities of the 2MFGC Galaxies

| 2MFGC | $l$ | $b$ | $J^c_{fe}$ | $W^c_{50}$ | $V_h$ | $V_{3K}$ | $H_0 r$ | $V_d$ | $V_{pec}$ | $V_p$ |
|---|---|---|---|---|---|---|---|---|---|---|
| 1 | 2 | | 3 | 4 | 5 | 6 | 7 | 8 | 9 | 10 |
| 2   | 91.89  | -65.99 | 12.15 | 389 | 6547  | 6196  | 6323  | -43  | -24   | 19 |
| 13  | 108.04 | -40.87 | 12.43 | 422 | 6804  | 6453  | 6704  | -125 | -135  | -10 |
| 16  | 102.30 | -54.33 | 12.40 | 595 | 14747 | 14386 | 13452 | -84  | 1402  | 1486 |
| 23  | 100.87 | -57.05 | 13.60 | 246 | 6339  | 5979  | 5521  | -75  | 537   | 612 |
| 31  | 110.45 | -34.25 | 13.64 | 280 | 7610  | 7272  | 6663  | -143 | 724   | 867 |
| 38  | 104.12 | -52.16 | 12.47 | 300 | 5538  | 5178  | 5530  | -92  | -273  | -181 |
| 49  | 320.05 | -64.94 | 12.97 | 404 | 10618 | 10426 | 10025 | 105  | 661   | 556 |
| 52  | 102.26 | -56.47 | 12.51 | 541 | 11563 | 11204 | 11512 | -78  | 34    | 112 |
| 53  | 109.09 | -40.63 | 9.42  | 383 | 2310  | 1961  | 2032  | -126 | -61   | 65 |
| 58  | 103.56 | -54.73 | 13.05 | 271 | 5071  | 4712  | 6115  | -84  | -1307 | -1223 |
| 59  | 103.41 | -55.06 | 13.10 | 207 | 3113  | 2754  | 4693  | -83  | -1883 | -1800 |
| 82  | 110.16 | -39.03 | 13.69 | 255 | 6597  | 6251  | 6885  | -131 | -511  | -380 |
| 88  | 332.05 | -73.41 | 10.66 | 291 | 1542  | 1304  | 2346  | 76   | -1028 | -1104 |
| 100 | 61.29  | -78.97 | 12.24 | 507 | 10178 | 9864  | 8766  | 10   | 1296  | 1287 |
| 105 | 112.95 | -29.18 | 10.98 | 454 | 5074  | 4751  | 5103  | -155 | -284  | -129 |
| 112 | 110.17 | -41.17 | 12.67 | 367 | 7843  | 7495  | 8361  | -126 | -686  | -560 |
| 122 | 114.23 | -24.47 | 11.06 | 432 | 4390  | 4082  | 4847  | -165 | -704  | -539 |
| 123 | 113.47 | -28.42 | 10.75 | 453 | 4811  | 4491  | 4533  | -157 | 11    | 168 |
| 127 | 358.00 | -79.76 | 11.92 | 444 | 6890  | 6615  | 7024  | 48   | -281  | -329 |
| 128 | 107.49 | -50.48 | 11.89 | 480 | 6674  | 6317  | 8912  | -99  | -2390 | -2291 |
| 136 | 112.68 | -33.67 | 12.44 | 377 | 8216  | 7883  | 7115  | -145 | 898   | 1044 |
| 137 | 322.31 | -69.51 | 10.76 | 466 | 6004  | 5791  | 4933  | 91   | 921   | 830 |
| 138 | 112.96 | -32.86 | 12.53 | 449 | 7037  | 6706  | 7950  | -147 | -1080 | -933 |
| 147 | 113.05 | -33.93 | 12.64 | 373 | 7445  | 7112  | 7841  | -145 | -570  | -425 |
| 155 | 106.69 | -55.94 | 12.68 | 348 | 8480  | 8124  | 7189  | -82  | 1069  | 1151 |
| 172 | 108.15 | -54.19 | 13.96 | 217 | 5836  | 5481  | 5304  | -88  | 249   | 337 |
| 177 | 111.80 | -43.27 | 10.70 | 562 | 5400  | 5052  | 5861  | -121 | -720  | -599 |
| 184 | 111.45 | -45.61 | 12.58 | 277 | 5620  | 5270  | 5505  | -115 | -157  | -43 |
| 190 | 111.55 | -45.77 | 11.55 | 308 | 4209  | 3859  | 3715  | -114 | 180   | 294 |
| 197 | 114.68 | -32.15 | 13.34 | 227 | 4863  | 4537  | 4995  | -150 | -394  | -244 |
| 198 | 109.07 | -54.65 | 10.92 | 400 | 3968  | 3614  | 4705  | -87  | -1034 | -946 |
| 205 | 116.71 | -20.05 | 12.59 | 347 | 5059  | 4769  | 7040  | -174 | -2142 | -1968 |
| 214 | 114.04 | -37.39 | 12.50 | 320 | 6047  | 5710  | 6260  | -137 | -449  | -312 |

The set of peculiar velocities $V_{pec}$ from Table 1 has been used to calculate the orthogonal components $\overline{V} = (V_x, V_y, V_z)$ of the dipole component of the bulk velocity,

$$V_{pec,i} = \overline{V} \cdot \overline{e}_i + V_{p,i} \qquad (5)$$

by minimizing the sum of the squares of the "noise" component $V_{p,i}$ of the peculiar velocity ($i$ is the sequence number of the galaxy in the sample). Here $\overline{e}_i = (\cos l_i \cos b_i, \sin l_i \cos b_i, \sin b_i)$ is the unit vector for the direction to the $i$-th galaxy in the data set associated with the galactic coordinates $l, b$. From the orthogonal components ($V_x, V_y, V_z$) we calculated the absolute value and direction of the bulk motion of the 2724 2MFGC galaxies: $V = 199 \pm 37$ km/s, $l = 304° \pm 11°$, and $b = -8° \pm 8°$ with $\sigma_{TF} = 0^m.47$ and $\sigma_V = 1018$ km/s. The errors in $V$, $l$, and $b$ were calculated in the following way: first we found the diagonal components $B_{VV}$, $B_{ll}$, and $B_{bb}$ of the covariation matrix **B** in the basis $\{\bar{e}_V, \bar{e}_l, \bar{e}_b\}$, and then evaluated the errors as $\Delta V = (B_{VV})^{1/2}$, $\Delta l = arctan\{(B_{ll})^{1/2}/V\}$, and $\Delta b = arctan\{(B_{bb})^{1/2}/V\}$.

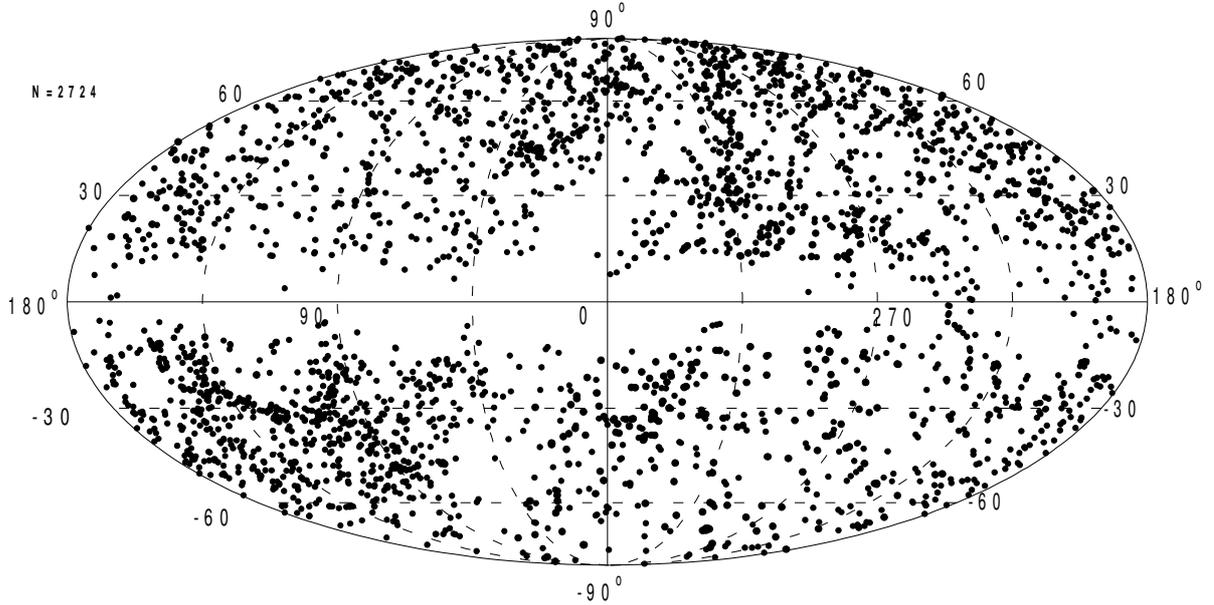

Fig. 3. The distribution of the N = 2724 2MFGC galaxies over the sky in galactic coordinates.

We emphasize that these results have been obtained for spiral galaxies of later types selected in terms of their infrared photometric characteristics in the XSC catalog. A comparison with Ref. 17 reveals an perfect mutual agreement in the calculated values of $V, l$, and $b$. Thus, the manner of selecting disk-shaped galaxies in terms of their optical features (RFGC) or their infrared characteristics (2MFGC) has little effect on the parameters of the dipole solution. The transition to a new sample ( RFGC $\rightarrow$ 2MFGC ) has made it possible to increase the number of galaxies in the sample to a depth of 19000 km/s by a factor of three. Because of this the accuracy of the determinations of $V, l$, and $b$ was increased by almost a factor of two, the significance of the dipole increased from 3.5 to 9.8 according to the Fisher criterion, and the "goodness" $G = (N/100)^{1/2}/\sigma_{TF}$ increased from 7.4 to 11.1.

Figure 4 shows the absolute value of the bulk velocity as a function of the maximum sample depth. This diagram shows that as the examined volume increases there is a substantial reduction in the amplitude of the bulk motion of the galaxies from 360 to 200 km/s. The damping of the amplitude of the non-Hubble motions with increasing scale (the convergence effect) is to be expected in the standard model for formation of the large scale structure of a universe with cold dark matter and a cosmological constant. Beyond $V_{3K} = 10000$ km/s the completeness of our sample falls off sharply, so the observed convergence effect actually corresponds to scales from $z = 0.01$ to 0.03.

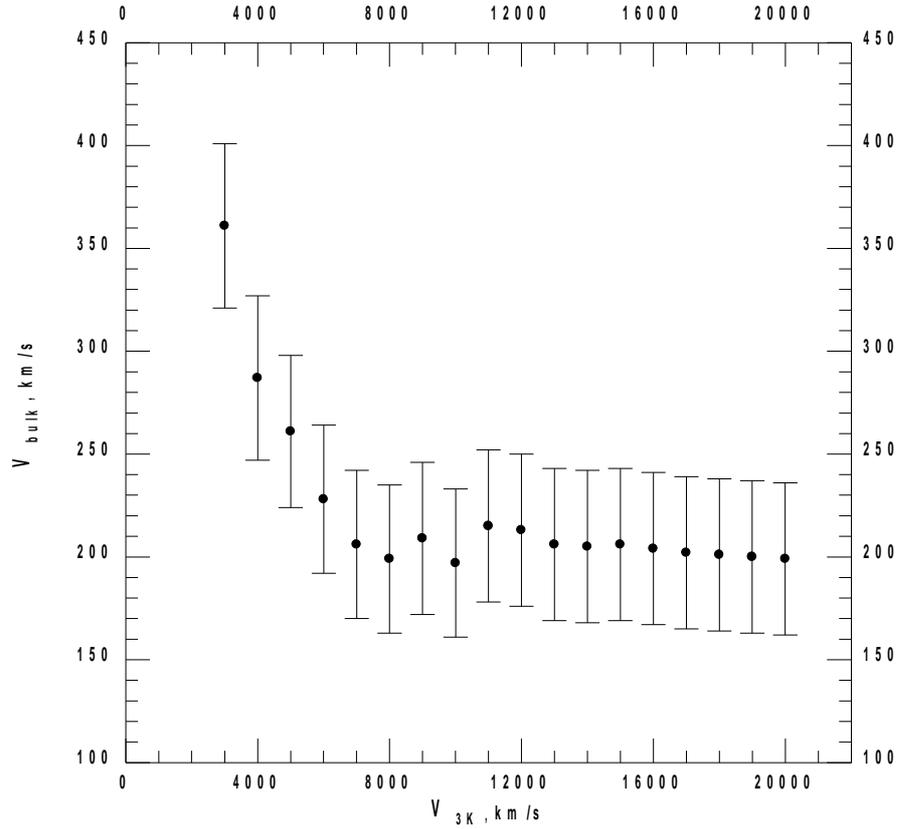

Fig. 4. The absolute value of the bulk velocity as a function of the maximum radial velocity of the galaxies in a given sample. The vertical bars indicate the standard error for the velocity.

## 4. Discussion of results

We now compare the present results with our previous estimates. In Section 2 we gave values of the parameters of the bulk motion for a cleaned sample of 971 flat spiral RFGC galaxies optically selected in photographic surveys of the sky. These parameters were obtained using the multi-parameter IRTF relation [17]. The results were: the velocity module $V = 199 \pm 61$ km/s, and apex coordinates $l = 301° \pm 18°$ and $b = -2° \pm 15°$, with $\sigma_{TF} = 0^m.42$ and $\sigma_V = 1045$ km/s. The maximum depth of the sample examined in this case was $V_{3K} = 18000$ km/s and the average $V_{3K} = 6160$ km/s.

Here we shall also give our earlier estimate of the parameters of the bulk motion for a cleaned sample of 919 RFGC galaxies with $V_{3K} < 18000$ km/s. From the optical parameters of these galaxies we obtained $V = 300 \pm 75$ km/s, $l = 328° \pm 15°$, and $b = +7° \pm 15°$, with $\sigma_V = 1160$ [6]. Here the values of $V, l,$ and $b$ were calculated based on our first list of peculiar velocities [5]. We can make a direct comparison of the samples, of the methods for estimating the distances, and of the individual distances obtained in Ref. 5, i.e., optical (OPT), and, in the present paper, infrared (IR).

Details regarding the sample construction, the description of the characteristics of the galaxies, the corrections applied to the observational data, and the methods for determining the distances are discussed in Ref. 6 and the references therein, as well as in the previous sections of this paper. Here we briefly list the features of the sets of data on the peculiar velocities of the galaxies.

**OPT. Sample of galaxies from the RFGC catalog.** The angular isophotal diameters are given in a uniform system that is close to the standard system $a_{25}$, and are corrected for the inclination and the galactic absorption in accordance with Ref. 29. The rotation amplitudes and widths $W_{20}$ are reduced to the line width $W_{50}$ according to the original sources and then corrected for cosmological broadening and turbulence. The heliocentric angular velocities are reduced to the $V_{3K}$ system. The individual distances $H_0 r$, meaning the distances by the angular isophotal diameter, were determined in one step (with simultaneous calibration of the TF relation and calculation of the distances) using the formula

$$H_0 r = [(c_1 + c_2 B + c_3 BT + c_4 a_{red}/a_{blue}) \cdot W_{50} + c_5] / a_{red} + c_6 (W_{50}/a_{red})^2 \qquad (6)$$

(Eq. (1) of Ref. 6), where $B$ is the optical surface brightness index, $T$ is the digitally expressed morphological type, $a_{red}$ and $a_{blue}$ are the angular diameters measured on the E and O Palomar prints. Here a regression was constructed for a subsample of 1132 galaxies which excluded galaxies with unreliable HI line profiles and distant galaxies with $z > 0.06$ from the original sample of 1327 RFGC galaxies. The peculiar velocities were defined as

$$V_p = V_{3K} - H_0 r - V_d, \qquad (7)$$

where $V_d$ is the dipole component of the radial velocity of the galaxies in the sample.

**IR. Sample of galaxies from the 2MFGC catalog.** Kron $J_{fe}$ values from the XSC catalog corrected for absorption in the galaxy. The rotation amplitudes and widths $W_{20}$ are reduced to $W_{50}$ using the relations found for this sample. The values of $W_{50}$ were corrected for cosmological broadening. The heliocentric radial velocities are reduced to the $V_{3K}$ system. The individual (luminosity) distances for the galaxies were calculated in two steps using a multi-parameter linear IRTF relation. Here relativistic corrections were incorporated in the relationships between the distance and redshift. The peculiar velocities $V_{pec}$ were determined using Eq. (4) of the present paper for the sample of 2724 2MFGC galaxies.

The distances for the 724 galaxies common to the OPT and IR lists are compared in Fig. 5.

It can be seen that, on the average, up to 12000 km/s the OPT and IR distances are related linearly. Different methods of processing the observational data (two- or one-step determination of the distance, use of IR magnitudes or optical diameter, etc.) lead to a random spread relative to a dependence that is linear on the average. But beginning roughly at 12000 km/s, a deviation from linearity can be seen. It is easy to show that the main reason for this nonlinearity is the presence in the model (Eq. (6)) of a quadratic term in the width $W_{50}$ which becomes substantial at large distances. Introducing this term at 3% reduces the dispersion in the "thermal" peculiar velocities (7), but its physical ground remains unclear. One of the reasons for the significance of this term in Eq. (6) according to the Fisher criterion may be the selection effect, whereby we choose larger galaxies at great distances.

We cannot say at present which of the two distances is more "correct." However, we can point out that the accuracy of the calculation of the absolute value of the bulk velocity is roughly twice as high for the IR sample than for the OPT sample (37 km/s vs. 75 km/s).

For studying the large scale motions of galaxies in a volume of radius ~100 Mpc we have placed a bet on thin, disk-shaped galaxies of later types which have a simple structure, are rich in gas, and, moreover, avoid the central regions of galactic clusters with their high virial velocities. Two catalogs, RFGC and 2MFGC, have been created for this purpose, with objects chosen on the basis of optical and infrared characteristics, respectively. Both catalogs cover the entire sky, both north and south. At present, only a small fraction of the galaxies in these samples (23% for RFGC and 15% for 2MFGC) have individual distance estimates made using the simple or multi-parameter Tully-Fisher relations. We believe that measurement of the rotation velocities and radial velocities of all the galaxies in both catalogs would provide an outline of the non-Hubble velocity field out to distances of ~200 Mpc with a very high detail.

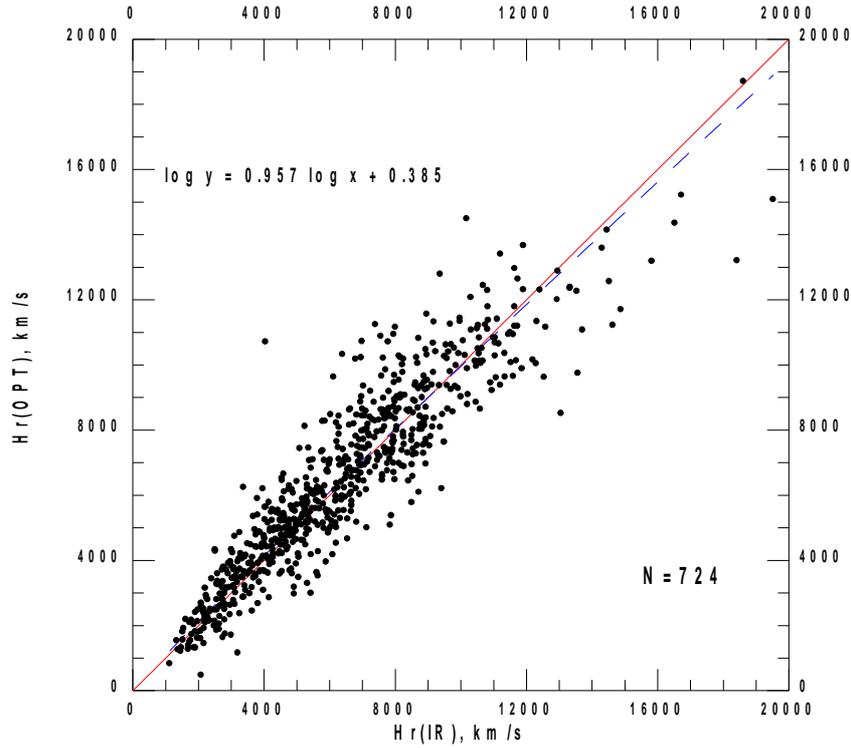

Fig. 5. The OPT distances as a function of the IR distances for a sample of 724 galaxies in common. The dashed line is a linear least squares regression for the logarithm of the distances. The regression equation is shown in the figure panel.

A comparison of the first list of peculiar velocities for flat spiral galaxies [5] with the list presented in this paper shows that the data are in good mutual agreement. The typical individual peculiar velocities in both samples are ~900-1000 km/s, owing primarily to errors in the measured distances to the galaxies. The distribution of peculiar velocities in the dipole approximation considered here has an amplitude of ~200-300 km/s in the $l \approx 301° \div 304°$ direction lying near the galactic plane with $b \approx -2° \div -8°$. The magnitude and direction of the dipole were essentially independent of whether the sample galaxies were selected for the optical or infrared characteristics, as well as of whether the optical angular diameters or the infrared stellar magnitudes were used in the Tully-Fisher relation. The stability of our determination of the dipole parameters and their consistence with the estimates of $V, l$, and $b$ based on the more representative samples by others [8-11] suggests that the value of the dipole vector at scales of ~100 Mpc has now been determined with an accuracy of up to ± 40 km/s and its direction, with an accuracy of ± 12°. A direct comparison of the individual distances for flat galaxies from their optical and infrared characteristics yields mutual agreement for velocities $V_{3K} \leq 12000$ km/s.

We have used data from the 2MASS survey and the NED and LEDA data bases in our work.